\documentclass[aps,pra,amsmath,amssymb,amsfonts,showpacs,superscriptaddress,floatfix,twocolumn]{revtex4} 
\usepackage{amsmath}
\usepackage{graphicx}
\usepackage{amsthm}
\usepackage{color}

\newcommand{\be}{\begin{equation}}
\newcommand{\ee}{\end{equation}}

\begin{document}
\title{Training A Quantum Optimizer}

\author{Dave Wecker}
\affiliation{Quantum Architectures and Computation Group, Microsoft Research, Redmond, WA 98052, USA}

\author{Matthew~B.~Hastings}
\affiliation{Station Q, Microsoft Research, Santa Barbara, CA 93106-6105, USA}
\affiliation{Quantum Architectures and Computation Group, Microsoft Research, Redmond, WA 98052, USA}

\author{Matthias Troyer}
\affiliation{Theoretische Physik, ETH Zurich, 8093 Zurich, Switzerland}
\affiliation{Quantum Architectures and Computation Group, Microsoft Research, Redmond, WA 98052, USA}
\affiliation{Station Q, Microsoft Research, Santa Barbara, CA 93106-6105, USA}
\begin{abstract}
We study a variant of the quantum approximate optimization algorithm [ E. Farhi, J. Goldstone, and S. Gutmann, arXiv:1411.4028] with slightly different parametrization and different objective: rather than looking for a state which approximately solves an optimization problem, our goal is to find a quantum algorithm that, given an instance of MAX-2-SAT, will produce a state with high overlap with the optimal state. Using a machine learning approach, we chose a ``training set" of instances and optimized the parameters to produce large overlap for the training set.  We then tested these optimized parameters on a larger instance set.  As a training set, we used a subset of the hard
instances studied by E. Crosson, E. Farhi, C. Yen-Yu Lin, H.-H. Lin, and P.
Shor (CFLLS) [arXiv:1401.7320]. When tested on the full set, the parameters that we find produce significantly larger overlap than the optimized annealing times of CFLLS.  Testing on other random instances from $20$ to $28$ bits continues to show improvement over annealing, with the improvement being most notable on the hardest instances.  Further tests on instances of MAX-3-SAT also showed improvement on the hardest instances.  This algorithm may be a possible application for near-term quantum computers with limited coherence times.
\end{abstract}
\maketitle
\section{Introduction}
The quantum approximation optimization algorithm (QAOA)\cite{QAOA,QAOA2} is a recently proposed quantum optimization algorithm, which itself is inspired by the quantum adiabatic algorithm (QAA)\cite{QAA}.  Consider a classical optimization problem.  Typically, the optimization problem will optimize some objective over bit strings of length $N$.  One encodes the objective function into
a quantum Hamiltonian $H_Z$ which is diagonal in the computational basis, using $N$ qubits to encode possible bit strings in the obvious way, with the optimal value of the objective function corresponding to the smallest value of $H_Z$.
Now define an additional Hamiltonian $H_X$, which is typically selected to be a transverse magnetic field on each qubit (the subscripts $X,Z$ on $H$ indicate whether the corresponding Hamiltonian is diagonal in the $Z$ basis or in the $X$ basis)

The QAA consists of first preparing the system in the ground state of Hamiltonian $H_X$ (which can be done easily since $H_X$ does not couple the different qubits) and then adiabatically evolving from $H_X$ to $H_Z$.  The simplest adiabatic path chosen is $H_s=(1-s)H_X+sH_Z$, for $s\in[0,1]$, although other paths have been considered\cite{dp}.  If the evolution time $T$ is sufficiently long compared to the smallest inverse spectral gap along the path (we denote the minimum gap as $\Delta_{min}$), then with probability close to 1 the final state will be the ground state of $H_Z$ and hence will solve the given instance.

Unfortunately, there are theoretical arguments that $\Delta_{min}$ can be super-exponentially small\cite{MBL} (scaling as $N^{-cN}$ for some constant $c>0$) for some instances, and so for these instances the time required for this adiabatic condition to hold is even longer than the time $2^N$ required by
an algorithm that iterates over spin configurations (other numerics suggests that the gap may not be quite as small as this for random instances\cite{apy}).
Some improvements have instead been found by looking at faster evolution times for which the adiabatic condition does not hold\cite{EC} and we review this in more detail below.

The QAOA is based on the observation that to implement the evolution under a time-dependent Hamiltonian on a quantum computer, the simplest method is
to Trotterize: first, decompose the evolution for a total time $T$ into many small increments ${\rm d}t$, small enough that the Hamiltonian $H_s$ is roughly
constant on time ${\rm d}T$.  Then, again for small enough ${\rm d}t$, one may decompose $\exp(i H_s {\rm d}t)\approx \exp(i (1-s) H_X {\rm d}t) \exp(i s H_Z {\rm d}t)$.  Thus, the total evolution is decomposed into a product of rotations by $H_X,H_Z$ with certain angles, and the final state at the end of the
evolution has the form
\begin{eqnarray}
\label{eq1}
 \Psi_F &=& \exp(i \theta_p^X H_X) \exp(i \theta_p^Z H_Z) \ldots \\ \nonumber && \exp(i \theta_2^X H_X) \exp(i \theta_2^Z H_Z) \exp(i \theta_1^X H_X) \exp(i \theta_1^Z H_Z) \Psi_I,
\end{eqnarray}
where $\theta_j^X,\theta_j^Z$ are some parameters determined by the evolution path, where the ``number of steps" $p=T/{\rm d}t$, and $\Psi_I$ is the ground state of $H_X$ (for all $j$, $\theta_j^X,\theta_j^Z$ are small, of order ${\rm d}t$, but for small $j$, $\theta_j^X$ is larger than $\theta_j^Z$ but for larger $j$ the reverse is true).
The QAOA then instead restricts to a much smaller value of $p$ (indeed, Refs.~\onlinecite{QAOA,QAOA2} study $p=1$) but allows the angles
$\theta_j^a$ to be chosen arbitrarily as variational parameters.  The parameters may then be adjusted to optimize some objective function; in Refs.~\onlinecite{QAOA,QAOA2}, this objective function was chosen to be the expectation value $\langle \Psi_F | H_Z | \Psi_F \rangle$.
 
In Ref.~\onlinecite{var}, a similar ansatz was used for purposes of approximating ground states of interacting quantum Hamiltonians, such as the Hubbard
model.  For example, in this case one might select $H_X$ to be a free fermion hopping term (or other term whose ground state can be easily prepared) and
$H_Z$ to contain the interactions.
Some modifications to the ansatz of Eq.~(\ref{eq1}) were made, as described in detail below.  A larger value of $p$ was chosen and a numerical search over parameter values was performed.

In this paper, we again use the modified ansatz of Ref.~\onlinecite{var}, but we apply it to the classical optimization problem of MAX-2-SAT.  Instead of adjusting parameters to minimize $\langle \Psi_F | H_Z | \Psi_F \rangle$, our objective function was the overlap between $\Psi_F$ and the true ground
state of the given instance.  We refer to this as ``targetting" the overlap.  Our general approach is inspired by machine learning techniques; this differs from the
 worst-case analysis of Refs.~\onlinecite{QAOA,QAOA2}.
We consider $p>1$ and we choose 
a ``training set" consisting of a small number of example instances.  This training set is chosen from the remarkable paper \cite{EC} which
searches for instances which are hard for the QAA and then investigates whether a fast anneal or other modifications outperforms the original algorithm.
After ``learning" a set of parameter values which optimize the average overlap on this
training set, we consider various test sets including many instances not in the training set.  We refer to a given sequence of parameters as a ``schedule". An ``annealing schedule" is a particular choice of parameters which approximates a linear anneal, so that the $\theta_j^X$ decrease linearly in $j$ while the $\theta_j^Z$ increase linearly in $j$, while a ``learned schedule" is a particular schedule obtained by optimizing parameters on a training set.

What  we find is that the schedules we have learned give results on various random test sets which outperform annealing schedules, including both slow and fast anneals (a sufficiently slow anneal will always find the ground state but for many of the test cases, the time required for such an anneal would be enormous, and if one restricts to anneals of modest time then a fast anneal outperforms a slow one).

Choosing a test set much larger than the training set is an essential step in showing the possible usefulness of this algorithm.  Learning a schedule
is very costly as it is done by a numerical search which itself consists of many steps and in each step we must evaluate the objective function, while testing the schedule requires a single evaluation of the objective function on each instance.

Further, we trained on sizes $N=20$ but tested on sizes up to $N=28$ where they continued to perform well and we also tested on some MAX-3-SAT instances.
All the simulations in this paper were performed on classical computers, taking a time exponential in $N$ and limiting the possible values of $N$.  However, if in the future a quantum computer becomes available, the algorithm could be run with larger values of $N$.  By training on a small size and testing on larger sizes, we raise the possibility that one might do training runs on a classical computer at smaller values of $N$ and then testing runs on a quantum computer at larger values of $N$ (one could also train on the quantum computer, of course, but time on the quantum computer may be more expensive than time on the classical computer; also, one might use the schedule found on the classical computer at small values of $N$ as a starting point for further optimization of the schedule at larger values of $N$ on the quantum computer).

\section{Problem Definition and Ansatz}
The MAX-2-SAT problem is defined as follows.  One has $N$ different Boolean variables, denoted $x_i$.  Clauses are made up from the  Boolean OR of two terms, each term being a variable or its negation.  Thus, possible clauses are all of one of the four forms
$$x_i \vee x_j, \; \overline{x}_{i} \vee x_j, \; x_i \vee \overline{x}_{j}, \; \overline{x}_{i} \vee \overline{x}_{j},$$
where $\overline{x}_{i}$ denotes the negation of a variable.  The problem is to find a choice of variables $x_i$ that maximizes the number of
satisfied clauses.

This problem can be cast into the form of an Ising model as follows.  Consider a system of $N$ qubits.  Let $\sigma^z_i$ denote the Pauli $Z$ operator on spin $i$.  Let $\sigma^z_i=+1$ correspond to $x_i$ being true and $\sigma^z_i=-1$ correspond to $x_i$ being false.
Then, a clause $x_i \vee x_j$
is true if
$\frac{1}{4}(1 - \sigma^z_i)(1 -\sigma^z_j)$
is equal to $0$ and is false if $\frac{1}{4}(1 -\sigma^z_i)(1-\sigma^z_j)=1$.
Indeed, each of the four possible types of clauses above can be encoded into a term
$$\frac{1}{4}(1\pm \sigma^z_i)(1\pm \sigma^z_j)$$ which is $0$ if the clause is true and $1$ if the clause is false, with the sign $\pm$ being chosen
based on whether the clause contains a variable or its negation.
Following CFLLS \cite{EC} which did an annealing study of the MAX-2-SAT problem, we define $H_Z$ to be the sum of these terms $\frac{1}{4}(1\pm \sigma^z_i)(1\pm \sigma^z_j)$ over all clauses in the instance.
Similarly following the notation of CFLLS, we define
\be
H_X=\sum_i \frac{1}{2}(1-\sigma^x_i),
\ee
where $\sigma^x_i$ is the Pauli $X$ operator on spin $i$.

With these choices of $H_X,H_Z$, the ground state energy of $H_X$ is equal to $0$ and the ground state energy of $H_Z$ is equal to the number of
violated clauses.  Both $H_X$ and $H_Z$ have integer eigenvalues.

As mentioned, Ref.~\onlinecite{var} used a modification of the ansatz (\ref{eq1}).  This ``modified ansatz" is
\begin{eqnarray}
\label{eq2}
 \Psi_F &=& \exp[i (\theta_p^X H_X +\theta_p^Z H_Z)] \ldots \\ \nonumber && \exp[i (\theta_2^X H_X+ \theta_2^Z H_Z)] \exp[i(\theta_1^X H_X+ \theta_1^Z H_Z)] \Psi_I.
\end{eqnarray}
The difference is that each exponential contains a sum of two non-commuting terms, both $H_X$ and $H_Z$.  We note that in the case of the
ansatz of Eq.~(\ref{eq1}), the quantities $\theta_j^a$ indeed are angles in that $\Psi_F$ is periodic in these quantities mod $2\pi$ if $H_X,H_Z$ have integer eigenvalues, but
for the modified ansatz of Eq.~(\ref{eq2}) the quantities $\theta_j^a$ are generally not periodic mod $2\pi$.
The modified ansatz was chosen because we found that
choosing the modified ansatz lead to a significantly easier numerical optimization in practice.
In the gate model of quantum computation, the simplest way to implement the modified ansatz is to approximate each exponential
$\exp[i(\theta_j^X H_X+\theta_j^Z H_Z)]$ using a Trotterization, which thus corresponds to a particular choice of parameters in the ``original ansatz" of
Eq.~(\ref{eq1}), albeit with a larger $p$.
In this paper we continue to use this ansatz.

\section{Training and Comparison To CFLLS}
\subsection{Problem Instances}
Our training sets are taken from examples in
CFLLS \cite{EC}.  We briefly review the construction of the instances there.  These are randomly constructed instances with $N=20$ variables and $60$ clauses. For each clause, the variables $i,j$ are chosen uniformly at random, and also each variable is equally likely to be negated or not negated, subject to the constraints that $i \neq j$ and that no clause appears twice, though the same pair of variables may appear in more than one clause.  Thus, it is permitted to have clauses $x_i \vee x_j$ and $x_i \vee \overline {x}_{j}$ but it is not permitted to have $x_i \vee x_j$ appear twice in the list of clauses.
From these random instances, further one retains only those instances that have a unique ground state.  In this way, $202,078$ instances were generated.
From these instances, a subset of hard instances are determined.  These are instances for which an implementation of the QAA using a linear annealing path $H_s=(1-s)H_X+sH_Z$ and an evolution time $T=100$ has a small success probability of less than $10^{-4}$ of finding the ground state.  In that paper, the Schr\"odinger equation was numerically integrated in continuous time.  This left a total of $137$ hard instances.  In the rest of the section, we simply call these ``instances", without specifying that they are
the hard instances.

For each instance, CFLLS then determined whether a {\it faster} anneal would lead 
to a higher probability of overlap with the ground state than the slow anneal of time $100$ (other strategies were considered as well in that paper, which we do not discuss here; we also remark that other authors have also considered the possibility of faster paths\cite{steiger,diab}).  The annealing time was optimized individually for each instance (keeping the annealing time smaller than $100$),
to maximize the squared overlap with the ground state \cite{pc}.
Below, when comparing learned schedules to annealing, we are comparing the ratio of the squared overlap for a learned schedule with that from this optimized anneal.
Our main result is that we are able to learn schedules for which this ratio is significantly larger than $1$.
If one instead made a comparison to a QAA with a fixed annealing time for all instances of CFLLS, this would lead to a further slight improvement in the ratio.

\subsection{Training Methods}
Rather than training on the full set of $137$ instances, we chose training sets consisting of $13$ randomly chosen instances from this set.  This was done partly to speed up the simulation, as then evaluating the average success probability could be done more rapidly on the smaller set, but it was primarily done so that then testing on the set of all instances would give a test set much larger than the training set; this is needed to determine whether the learned parameters
generalize to other instances beyond the training set.

Given a training set, our objective function is the average, over the training set, of the squared overlap between the state $\Psi_F$ and the ground state
of $H_Z$.
To compute the objective function, we compute the state $\Psi_F$; we do this by approximating the exponentials $ \exp[i(\theta_j^X H_X+ \theta_j^Z H_Z)]$ by a Trotter-Suzuki formula, as 
\begin{eqnarray}
&&\exp[i(\theta_j^X H_X+ \theta_j^Z H_Z)] \nonumber \\ \nonumber &\approx & \Bigl(
\exp(i \frac{\theta_j^Z}{2n} H_Z)
\exp(i \frac{\theta_j^X}{n} H_X)
\exp(i \frac{\theta_j^Z}{2n} H_Z) \Bigr)^n,\end{eqnarray}
where we chose $n=4$. 
This value of $n$ was chosen as the smallest value of $n$ that gives results for an annealing schedule on the CFLLS data set which are consistent with the continuous time limit; larger values of $n$ will likely lead to slight changes in the optimal parameters of the learned schedule.

We treat this objective function as a black box, and optimize the parameters in the schedule using the same algorithm as in Ref.~\onlinecite{var}, except for
modification of how we choose the starting point for the search (also, we do {\it not} use the annealed variational method of Ref.~\onlinecite{var} to do the search).
Briefly, the optimization algorithm is: given an ``initial schedule" (i.e., a schedule chosen as the starting point for the optiization), we use a greedy noisy search, slightly perturbing the values of each $\theta_j^a$ at random, accepting the perturbation if it improves the objective function for a total of $150$ evaluations of
the objective function.  The step size for the greedy search is determined in a simple way: every fifty trials, we count the number of acceptances.  If the number is large, the step size is increased and if the number is small the step size is reduced\cite{detail}.  After the noisy search, we then use Powell's conjugate direction\cite{nrep} method until it converges.  We alternate Powell's method and the noisy search until no further improvement is obtained. 

We did this numerical optimization for $5$ different randomly chosen training sets of 13 instances (10\% of the data for each).  For each training set, we did $5$ different runs of the optimization for
a variety of initial schedules, thus giving $25$ runs for each initial schedule.  While different choices of initial schedule led to very different performances of the final schedule found at the end of the optimization, for any given choice of initial schedule the results were roughly consistent across different
choices of the training set and different optimization runs.  Certain training sets tended to do slightly better (schedules trained on them tended to perform better when tested on the full set as described in the next section) but in general for an appropriate choice of initial schedules we found that
{\it all} choices of training sets and {\it all} runs of the optimization with that initial schedule and training set led to good performance on the full set.

\subsection{Results}
\begin{figure}
\includegraphics[width=3.5in]{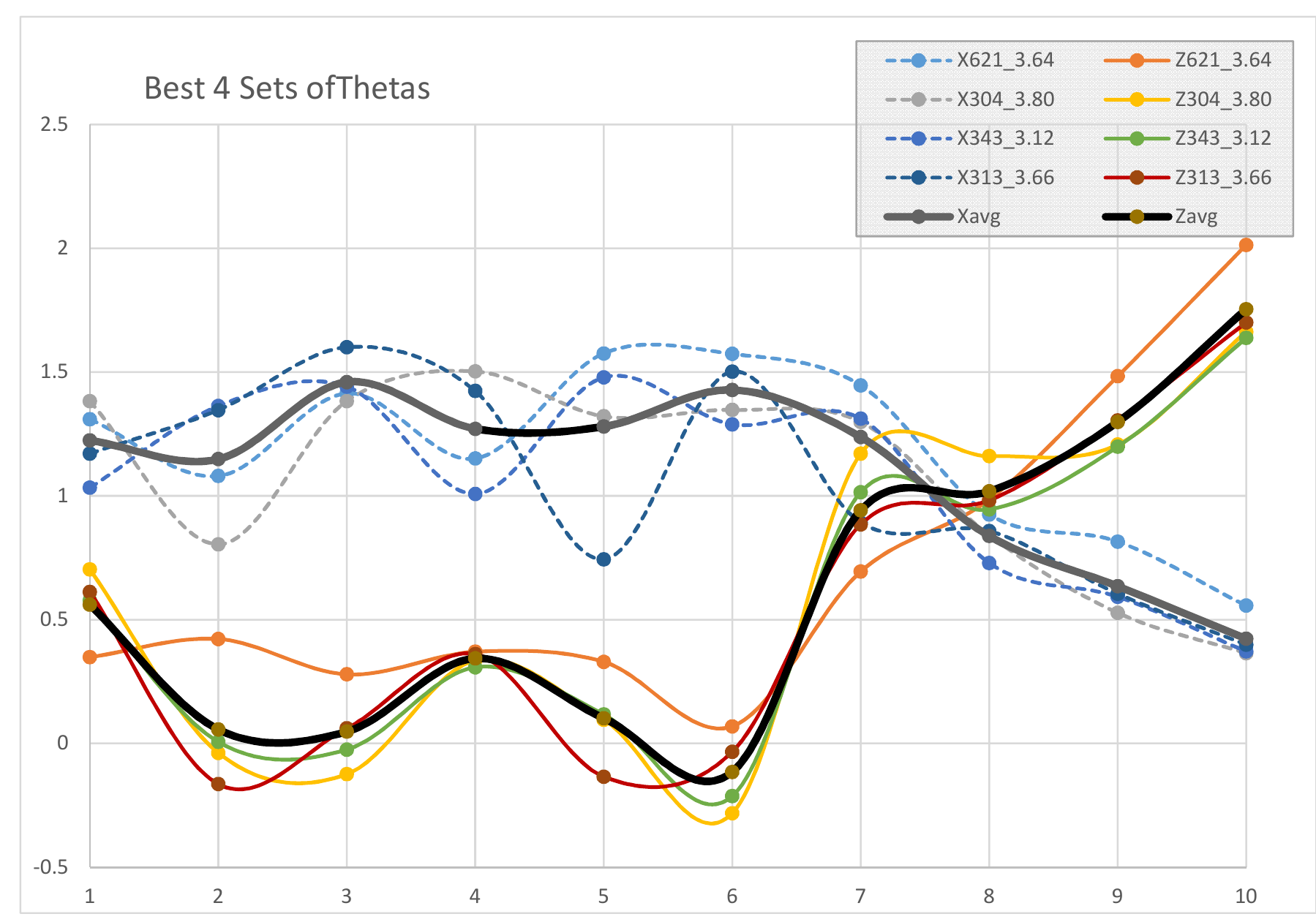}
\caption{Dashed curves show $\theta^X$ and solid curves show $\theta^Z$.  Four different learned schedules are shown; the format such as $X621\_3.64$ indicates that this curve is $\theta^X$, for a schedule started using initial schedule $6$; the $21$ indicate the particular training set and run (these numbers are not important, as they are just keys to a random number generator but they differentiate the three different curves that use initial schedule $3$); the $3.64$ indicates the average improvement for that schedule.  The $Xavg,Zavg$ curves show the parameters averaged over those four schedules.}
\label{figsched}
\end{figure}

The learned schedules that performed well had a form quite different from an annealing schedule.  Instead, the form of many of the good schedules was similar to that in
Fig.~\ref{figsched}.
The schedule begins with $\theta^X$ large and fairly flat but $\theta^Z$ oscillating near zero.  Then, at the end of the schedule, the values are more reminiscent of an anneal, with $\theta^Z$ increasing (albeit with some oscillations) and $\theta^X$ decreasing fairly linearly.

To find the schedules shown in Fig.~\ref{figsched} required an appropriate choice of initial schedule (described further below).
Instead, if we chose an initial schedule that was an annealing schedule, the search over schedules would become stuck in local optima that did not perform as well.

After discovering this form after some experimentation, we studied a variety of schedules which had this form.  These schedules were labelled by a key ranging from $2$ to $14$ (key values of $0,1$ corresponded to schedules with a different form that did not perform well and are not reported here).
These schedules are shown in Table \ref{tabsched}.

The details of the schedules are not that important.  We simply report the variety of the schedules considered for completeness and to show that all such choices led to some improvement but that certain choices consistently led to more improvement.  Some of the schedules are described as ``Frozen"; in this case, the $\theta^Z$ variables were not allowed to change during the learning process and only the $\theta^X$ variables were allowed to change.  Thus, the final learned schedule had the same $\theta^Z$ variables as the initial and this was chosen to be
$\theta_j^Z$ changing linearly as a function of $j$.  These schedules may be simpler to implement in hardware due to less need for complicated control of $\theta^Z$.
They showed some improvement but not quite as much as others.

\begin{table}
\begin{tabular}{c|c|c}
Key & $\theta^X$ & $\theta^Z$ \\
2 & 1111111111 & 0000000000 \\
3 & 1111111110 & 0000000001 \\
4 & 1111100000 & 0000011111 \\
5 & 0000000000 & 1111111111 \\
6 & 1111111111 & Linear \\
7 & 1111100000 & Linear \\
8 & 1111111110 & 000000001 \\
9 & 1111111110 & Linear \\
10 & 1111111110 & Frozen \\
11 & 1111111150 & 0000000051 \\
12 & 1111111150 & Linear \\
13 & 1111111150 & Frozen \\
14 & Avg & Avg 
\end{tabular}
\caption{Initial schedules for $\theta^Z_j,\theta^X_j$.  The $10$ entries in a line such as ``1111111150" shows a sequences of $\theta_j$ for $j=1,...,10$ in order.
An entry $1$ or $0$ indicates a $1$ or $0$, while $5$ indicates $0.5$.  ``Linear" indicates a linear function, $\theta^Z_j=0.05,0.15,...,0.95$ for $j=1,...,10$.
``Frozen" also indicates a linear function, but with $\theta^Z$ held fixed during learning as described in text.  ``Avg" indicates that the initial schedule is the average schedule shown in
Fig.~\ref{figsched}.}
\label{tabsched}
\end{table}

The improvement is shown in Table \ref{tabimp}.  The data in this table includes all $137$ instances, so it includes instances which are in the
training set; however, these instances represent less than $10\%$ of the test set.  We report in this table a ``ratio of averages".  That is, we compute the squared overlap of $\Psi_F$ with the ground state for each intance and average over instances.  Then, we compute the ratio of this average to the same average using the
optimized annealing times of CFLLS.
The parameters for certain schedules which performed well are shown in the Appendix.

\begin{table}
\begin{tabular}{c|c|c|c|c|c|c}
Initial & 0 & 1 & 2 & 3 & 4 &Avg \\
\hline
2 & 1.4 & 2.0 & 1.7 & 1.9 & 2.2 & 1.8\\
3 & 4.2 & 3.6 & 3.5 & 3.3 & 3.8 & 3.7\\
4 & 2.5 & 2.4 & 2.4 & 2.3 & 2.4 & 2.4\\
5 & 2.4 & 2.3 & 2.4 & 2.4 & 2.4 & 2.4\\
6 & 2.9 & 3.0 & 3.1 & 3.3 & 2.6 & 3.0\\
7 & 2.4 & 2.0 & 2.3 & 2.2 & 2.1 & 2.2\\
8 & 3.5 & 3.5 & 3.4 & 3.5 & 3.7 & 3.5\\
9 & 2.7 & 3.2 & 2.8 & 3.1 & 3.4 & 3.0\\
10 & 2.5 & 2.2 & 2.1 & 2.4 & 2.1 & 2.3\\
11 & 4.4 & 4.2 & 4.2 & 4.1 & 4.1 & 4.2\\
12 & 3.1 & 2.9 & 3.3 & 3.5 & 3.1 & 3.2\\
13 & 2.0 & 2.4 & 2.3 & 2.0 & 2.0 & 2.1\\
14 & 4.5 & 4.5 & 4.3 & 4.5 & 4.4 & 4.4\\
\hline
Avg & 3.0 & 2.9 & 2.9 & 2.9 & 3.0 & 2.9
\end{tabular}
\caption{Improvement compared to optimized annealing times.  The entries report the ratio of averages (see text).
First column ``Initial" labels the initial schedule from table \ref{tabsched}.  Columns $0,1,2,3,4$ label different training sets.  Column ``Avg" is average of that row over training sets.  Row ``Avg" is average of that
training set over choices of Initial.  One can see that there is some variance from one training set to another, but the performance is roughly consistent.
The best rows are 14, 11 and 8. }
\label{tabimp}
\end{table}

\begin{figure}
\includegraphics[width=3.5in]{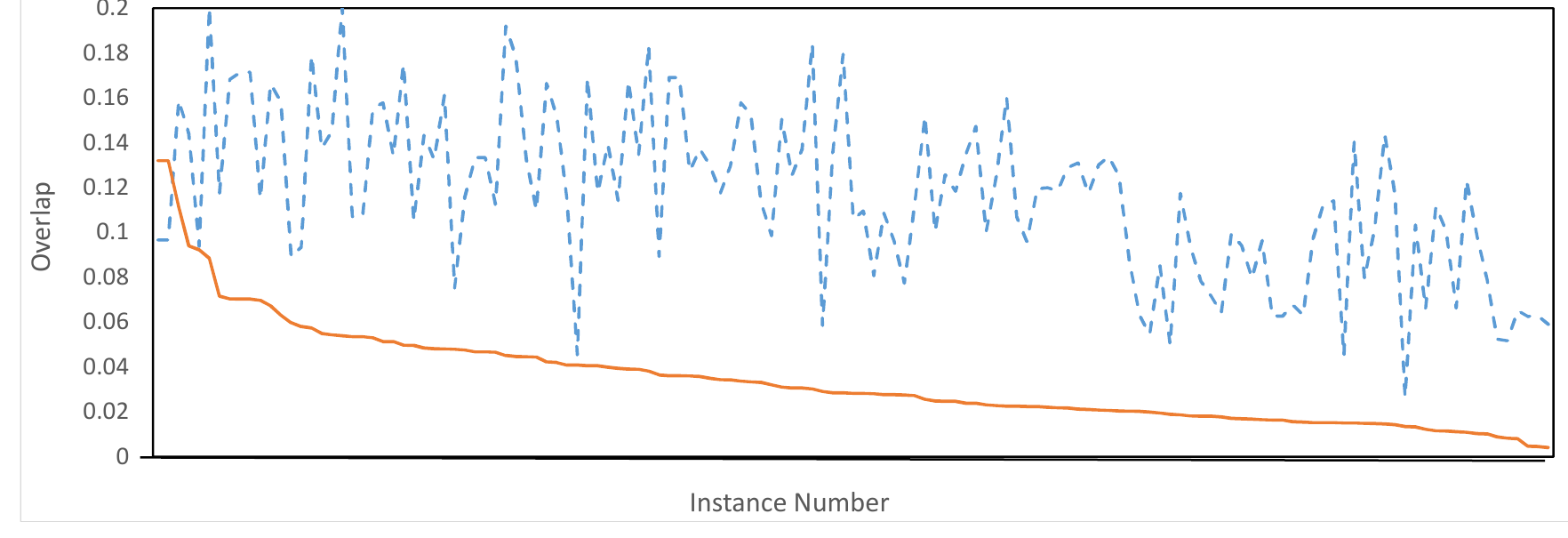}
\caption{$x$-axis labels different instances.  $y$-axis shows overlap.  Dashed curve is from learned schedule while solid curve is for optimized anneal.  Instance numbers differ from CFLLS because instances are sorted by overlap for optimized anneal.}
\label{figh}
\end{figure}

Another option to reporting the ``ratio of averages" is to report an ``average of ratios".  This means computing, for each instance, the ratio of the squared overlap of $\Psi_F$ with the ground state for a given learned schedule to the same overlap for an optimized anneal.  Then, averaging this ratio over intances.
The result would be different and would lead to a larger improvement because the learned schedules do better on the harder instances as shown
in Fig.~\ref{figh}.

\section{Testing on Random Instances with $N=20,24$ and $28$}
In addition to testing against the instances of CFLLS, to determine whether the learned schedules generalize to larger sizes and other ensembles, we constructed further problem instances for $N=20,24$ and $28$. 
We repeated the case $N=20$, since the ensemble that we constructed differs from that in CFLLS as we explain.

We took $60,72$ and $84$ clauses, respectively, so that the clause-to-variable ratio was maintained.  We used the same ensemble as in CFLLS, so 
that clauses are chosen at random subject to the constraint that no clause appears twice and that the instance has a unique ground state.
However, rather than finding hard instances based on a continuous time anneal at time $T=100$, we used a slightly different method .
This was partly done to speed up our search for hard instances; in CFLLS, fewer than $1/1000$ of the instances were hard by that standard.
However, it was primarily done to test the learned schedules in a more general setting and to consider a range of hardnesses to demonstrate that
the learned schedules perform relatively better on the harder instances.

In testing hardness, we used annealing schedules.  Since we will compare to a variety of annealing schedules, we introduce some notation.  Let
$L(p,x,z)$ denote the schedule with $p$ steps and $\theta_j^Z=zj/(p+1)$ and $\theta_j^X=x(p+1-j)/(p+1)$.

We used $L(10,1,1)$ to determine hardness, constructing 3346 random instances and sampling from $6.8\%$ of the instances which had the smallest
squared overlap with $L(10,1,1)$, yielding 170 instances (for $N=28$, we generated a smaller number of instances so that only $72$ were retained).
The reason for choosing $6.8\%$ is that the resulting ensemble had a difficulty for $L(10,1,1)$ which was roughly comparable to that of the CFLLS instances (however, the actual distribution of instance difficulty is different from CFLLS and so the value $6.8\%$ is fairly arbitrary).
On these instances, a comparison of various algorithms is shown in Tables \ref{tabar} and \ref{tabra}.
We also include in these tables results for the instances of CFLLS, as now the tables compare the performance of various learned schedules to
$L(10,1,1)$ rather than to an optimized anneal.
For the instances described in this section, we only compared to schedules of the form $L(p,x,z)$ which give a discrete approximation to an anneal, rather than comparing to anneal.
This was done to simplify the numerics.  The results for the instances of CFLLS is that such schedules give performance similar to that of a continuous time QAA.

In these tables, the learned schedules are identified by a pair such as $31 (9)$.  In this case, the number $31$ is an arbitrary key labelling the schedule.  The number in parenthesis, $9$ in this case, indicates that schedule $31$ was obtained by starting from initial schedule $9$ in Table \ref{tabsched}.
We only give the keys here because we also later refer to certain schedules by key; in particular, number 154 which is one of the best performing by several measures.

\begin{table}
\scalebox{0.9}{
\begin{tabular}{c||c|c|c|c||c|c|c|c}
& Overlap &&&& Ratio &&&\\
\hline
Sched & CFLLS & N=20 & N=24 & N=28 & CFLLS & N=20 & N=24 & N=28\\
\hline
8 (8)& 0.111 & 0.068 & 0.040 & 0.025 & 11.9 & 4.4 & 6.7 & 8.2\\
31 (9)& 0.108 & 0.048 & 0.028 & 0.017 & 8.1 & 2.9 & 4.0 & 5.0\\
49 (9)& 0.108 & 0.026 & 0.013 & 0.007 & 6.6 & 1.6 & 1.7 & 2.0\\
84 (11)& 0.120 & 0.065 & 0.037 & 0.023 & 10.4 & 4.1 & 5.9 & 7.1\\
113 (12)& 0.111 & 0.024 & 0.011 & 0.006 & 6.8 & 1.5 & 1.6 & 1.8\\
122 (12)& 0.107 & 0.029 & 0.014 & 0.008 & 7.0 & 1.7 & 1.9 & 2.3\\
154 (14)& 0.117 & 0.085 & 0.050 & 0.034 & 10.5 & 5.2 & 7.7 & 10.5\\
157 (14)& 0.116 & 0.079 & 0.047 & 0.032 & 10.6 & 4.9 & 7.4 & 9.8\\
L(10,1,1) & 0.025 & 0.019 & 0.009 & 0.004 & 1.0 & 1.0 & 1.0 & 1.0\\
L(10,2,2) & 0.024 & 0.075 & 0.039 & 0.021 & 1.0 & 4.0 & 5.1 & 5.3\\
L(10,3,3) & 0.011 & 0.105 & 0.058 & 0.032 & 0.5 & 5.8 & 8.3 & 8.3\\
L(10,4,4) & 0.006 & 0.118 & 0.056 & 0.038 & 0.3 & 6.5 & 13.5 & 9.7\\
L(20,1,1) & 0.028 & 0.073 & 0.028 & 0.022 & 1.3 & 3.9 & 6.4 & 5.4\\
L(40,1,1) & 0.008 & 0.159 & 0.077 & 0.054 & 0.4 & 8.8 & 18.8 & 14.1\\
L(80,1,1)& 0.0003 & 0.288 & 0.164 & 0.132 & 0.0 & 16.3 & 43.5 & 34.1
\end{tabular}}
\caption{First column labels schedule.  Next four columns
gives the average overlap for various test sets for each schedule; $N=20,24$ and $28$ refers to random instances constructed following procedure described in this section.  Last four columns
give average (over instances) of ratio (of square overlap) comparing to $L(10,1,1)$.  Note that the entry in the last four columns is $1$ for the
schedule $L(10,1,1)$ because there it is being compare to itself.}
\label{tabar}
\end{table}

\begin{table}
\scalebox{0.9}{
\begin{tabular}{c||c|c|c|c}
& Ratio &&&\\
\hline
Sched & CFLLS & N=20 & N=24 & N=28\\
\hline
8 & 4.4 & 3.5 & 4.7 & 5.6\\
31 & 4.2 & 2.5 & 3.2 & 3.7\\
49 & 4.2 & 1.4 & 1.5 & 1.6\\
84 & 4.7 & 3.4 & 4.3 & 5.3\\
113 & 4.4 & 1.3 & 1.3 & 1.4\\
122 & 4.2 & 1.5 & 1.7 & 1.8\\
154 & 4.6 & 4.4 & 5.9 & 7.6\\
157 & 4.6 & 4.1 & 5.5 & 7.1\\
L(10,1,1) & 1.0 & 1.0 & 1.0 & 1.0\\
L(10,2,2) & 0.9 & 3.9 & 4.5 & 4.8\\
L(10,3,3) & 0.4 & 5.5 & 6.8 & 7.2\\
L(10,4,4) & 0.2 & 6.2 & 6.5 & 8.5\\
L(20,1,1) & 1.1 & 3.8 & 3.3 & 4.9\\
L(40,1,1) & 0.3 & 8.3 & 9.0 & 12.3\\
L(80,1,1) & 0.01 & 15.0 & 19.2 & 29.8
\end{tabular}}
\caption{First column labels schedule.  Next four columns
give ratio of average comparing to $L(10,1,1)$ for various test sets.  Note that the entry in the last four columns is $1$ for the
schedule $L(10,1,1)$ because there it is being compare to itself.}
\label{tabra}
\end{table}

Note that while the learned schedules, in particular $154$, improve over $L(10,1,1)$, we find that slower anneals such as $L(80,1,1)$ outperform the learned
schedules {\it on the $N=20,24$ and $28$ instances}.  However, on instances from CFLLS, the slower annealing schedules do significantly worse, with
$L(80,1,1)$ much worse than $L(10,1,1)$.

\begin{figure}
\includegraphics[width=3in]{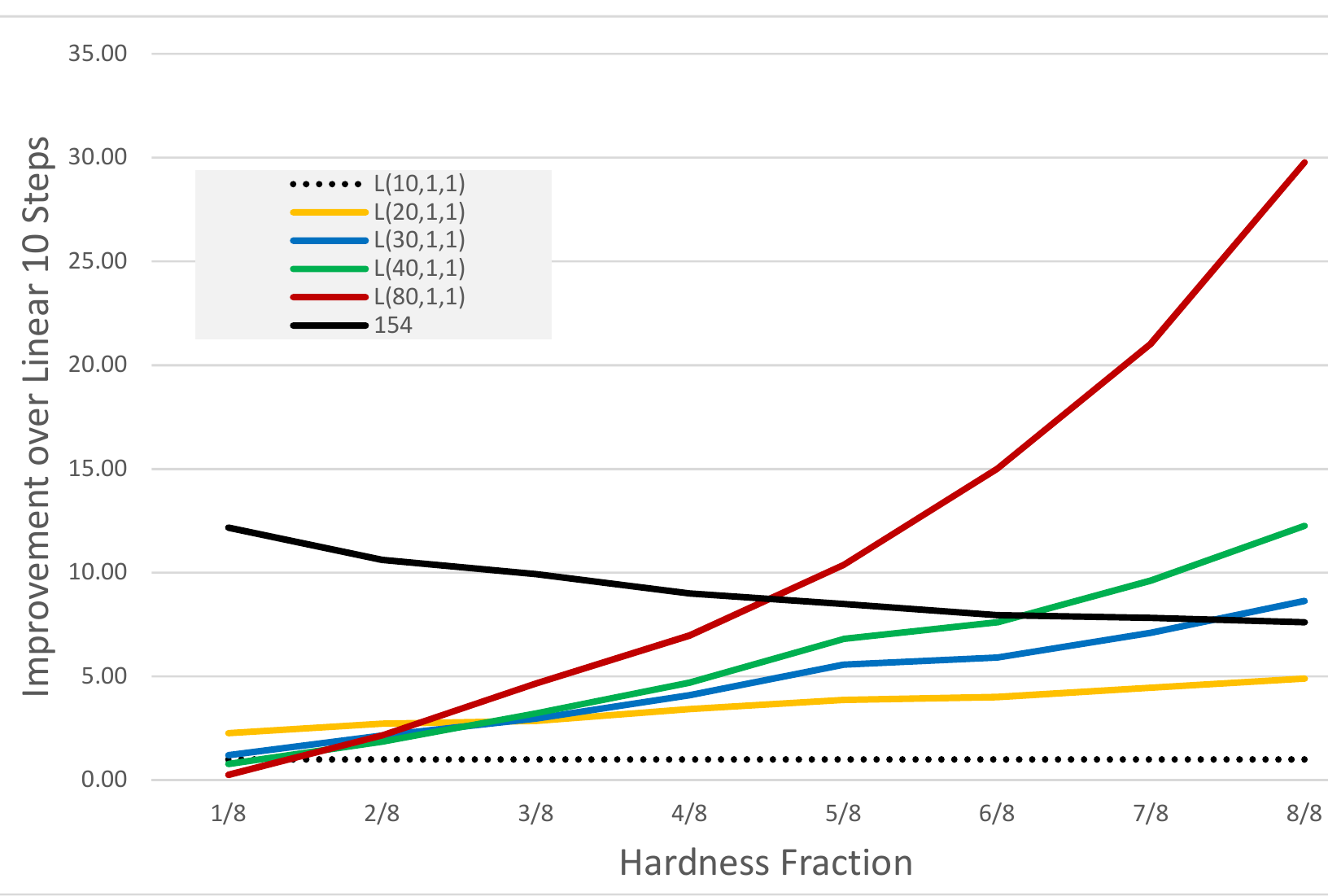}
\caption{Ratio of averages for $8$ different subsets of the MAX-2-SAT instances with $N=20$, chosen by binning by hardness for $L(80,1,1)$.  We compare various schedules to $L(10,1,1)$.
Different colors label different schedules.  On hardest instances, $154$ does best, followed by $L(20,1,1),L(30,1,1),L(10,1,1),L(40,1,1),L(80,1,1)$ in sequence. Hardest instances are on the left of the graph.}
\label{figbin}
\end{figure}

The reason for this can be seen by further dividing the instances based on their hardness for $L(80,1,1)$.  We binned the instances into $8$ different groups depending upon the squared overlap for $L(80,1,1)$.  Fig.~\ref{figbin} shows the performance compared to $L(10,1,1)$ of various schedules for 
each bin.  We find that learned schedule $154$ (chosen simply as it was the best example, we expect similar performance from other learned schedules) outperform $L(10,1,1)$ everywhere, while  the performance compared to $L(80,1,1)$ varies: it outperforms $L(80,1,1)$ on the instances where $L(80,1,1)$ does worst.  On the instances where $L(80,1,1)$ does worst, even $L(10,1,1)$ outperforms $L(80,1,1)$.  This fits with the observed performance of the learned schedule on the instances of CFLLS as those instances were chosen to be difficult for a slow anneal.

Importantly, the data shows that as $N$ increases the ratio between the learned schedules and $L(10,1,1)$ is {\it increasing}.
This may partly be due to the fact that the overlap for all schedules is decreasing with increasing $N$.

\subsection{MAX-3-SAT}
\begin{figure}
\includegraphics[width=3in]{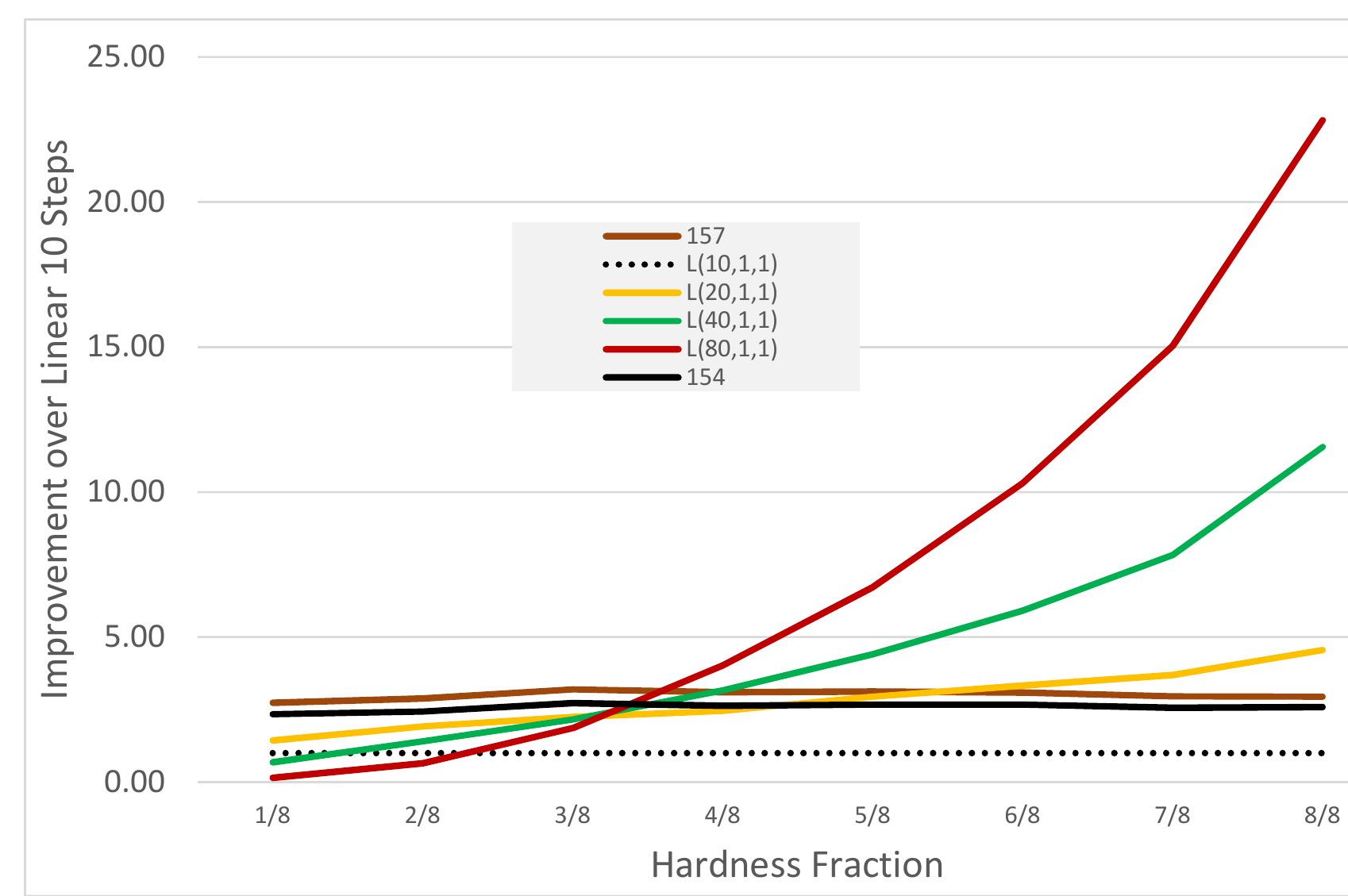}
\caption{Ratio of averages for $8$ different subsets of the MAX-3-SAT instances with $N=20$, chosen by binning by hardness for $L(80,1,1)$.  We compare various schedules to $L(10,1,1)$.  $154$ and $157$ are both learned schedules.
Different colors label different schedules.  On hardest instances, $157$ does best, followed by $154,L(20,1,1),L(10,1,1),L(40,1,1),L(80,1,1)$ in sequence.}
\label{figbin3}
\end{figure}

As a final example, we tested the performance of the algorithm on MAX-3-SAT.  Clauses were of the form $x_i \vee x_j \vee x_k$ (or similar, with some variables negated).  Each variable in the clause was chosen independently and uniformly and was equally likely to be negated or not negated (so in this case it is possible to have a clause such as $x_i \vee x_i \vee x_j$ which is just a 2-SAT clause or a clause such as $x_i \vee \overline{x}_{i} \vee x_j$ which is always true).  We took $N=20$ variables and $120$ clauses (clauses were chosen independently and we allowed the same clause to occur more than once).
The clause to variable ratio was taken $6$ to ensure that we are above the satisfiability phase transition\cite{3satpt}.
We then selected for intances which had unique ground states.
Finally we chose the hardest $6.8\%$ of instances based on overlap for $L(10,1,1)$.  The results are shown in Fig.~\ref{figbin3}.
We emphasize that we use the schedules trained on MAX-2-SAT instances from CFLLS here, even though this is a different problem.

\section{Toy Model and Theoretical Analysis}
\subsection{Toy Model}
To better understand {\it why} the learned schedules perform well, we have constructed a toy model.  We write the model directly as an Ising model (it does not exactly correspond to a MAX-2-SAT instance since some of the terms involve only a single variable).
The model is related to a model studied in Refs.~\onlinecite{ring1,ring2} but with one crucial modification; in those papers, a model was studied which
has a large number of classical ground states.  All but one of those ground states form a cluster of solutions which are connected by single spin flips, while the remaining ground state is isolated from the others and can only be reached by flipping a large number of spins.  It was shown that a quantum annealer will be very likely to end at one of the ground states in the cluster, while a classical annealer in contrast will have a much higher probability of ending at the isolated ground state.  We modify this model so that it has only a single unique ground state (the isolated state of the original model), moving the others to
higher energy.  In this way, it becomes very difficult for a quantum annealer to locate the ground state.
\begin{figure}[tb]
\includegraphics[width=0.7\columnwidth]{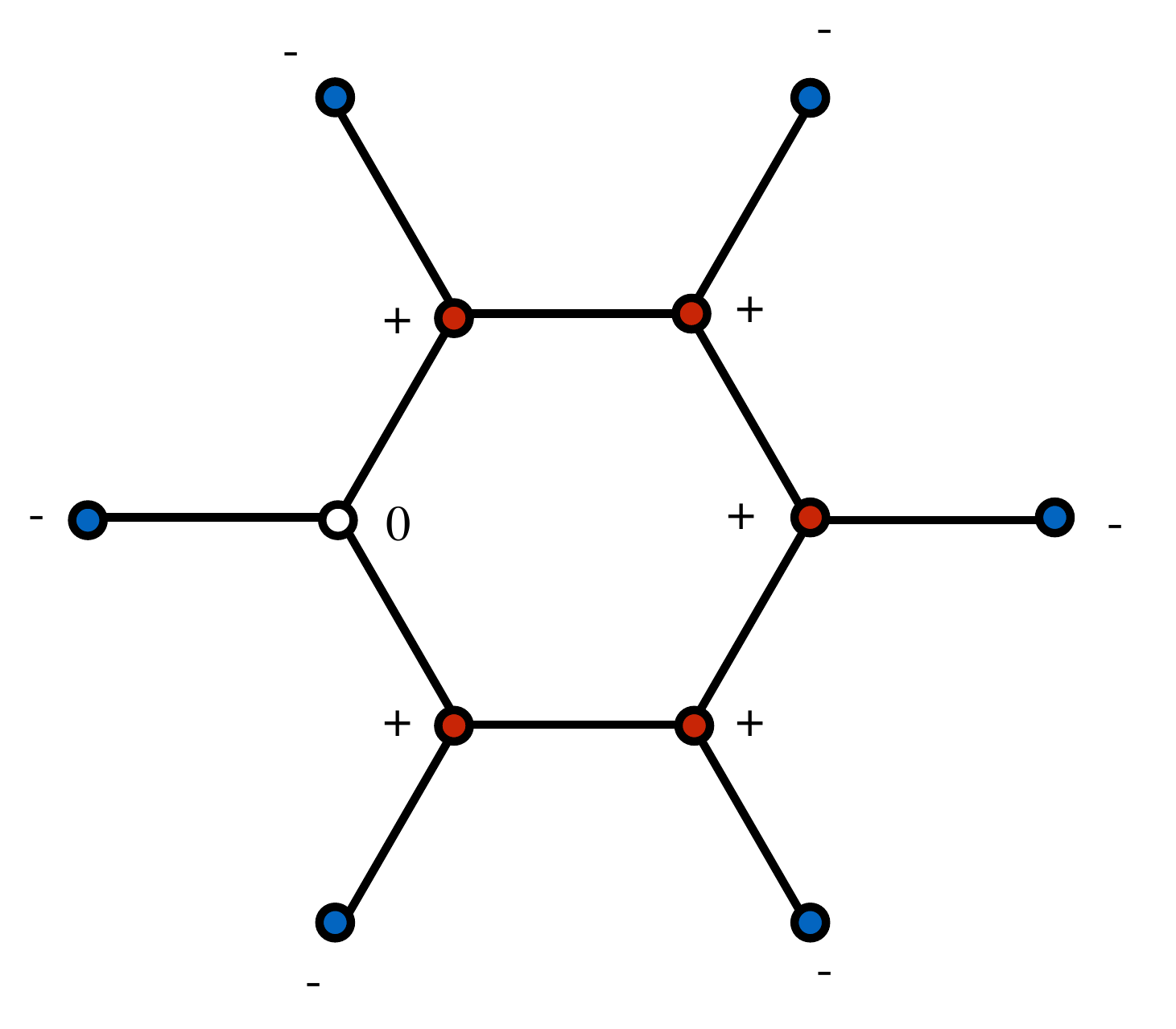}
\caption{Graph of the toy model considered for the case of $K=6$ with $N=12$ spins. The edges indicate ferromagnetic couplings between spins. All but one spin of the inner ring has positive magnetic fields (indicated by $+$ symbols), while all the outer spins have negative fields (indicated by $-$) symbols applied in the $z$-direction.}
\label{fig:graph} 
\end{figure}

This is a model with $N=2K$ spins.  As shown in Fig. \ref{fig:graph}, $K$ of the spins form what is called the ``inner ring", and are arranged in a ring with ferromagnetic couplings
of strength $1/4$.  The $1/4$ is chosen to correspond to the factor of $1/4$ that arises when translating from a MAX-2-SAT model to an Ising model; we chose
to keep the magnitudes of terms similar to the magnitudes of the terms on the training set.
Each of the other spins form what is called the ``outer ring".  The outer ring spins are not coupled to each other; instead, each outer ring spin is coupled to one inner ring spin (every outer ring spin is coupled to a different inner ring spin), again with ferromagnetic couplings of strength $1/4$.
Finally, on every outer ring spin there is a magnetic field in the $Z$ direction with strength $-1/4$ while on all but one of the the inner ring spins,
there is a $Z$ direction magnetic field with strength $+1/4$.  
Thus, labelling the spins by $i=0,\ldots,N-1$ with $0\leq i <K$ corresponding to the inner ring, we have
\begin{eqnarray}
\label{toy}
H_Z&=&-\frac{1}{4}\sum_{i=0}^{K-1} S^z_i S^z_{i +1 {\rm mod} \, K}-\frac{1}{4}\sum_{i=0}^{K-1} S^z_i S^z_{i+K} \\ \nonumber && -\frac{1}{4}\sum_{i=K}^{2K-1} S^z_i
+\frac{1}{4}\sum_{i=0}^{K-2} S^z_i.
\end{eqnarray}

To better understand this model, suppose that instead we added the $Z$ direction magnetic field with strength $+1/4$ to {\it all} spins on the inner
ring, so that the last term of $H_Z$ became
$\frac{1}{4}\sum_{i=0}^{K-1} S^z_i$.  This model, which is the model studied in Refs.~\onlinecite{ring1,ring2}, has $2^K+1$ degenerate ground states.
The isolated ground state is the state with $S^z_i=+1$ for all $i$.  The cluster of $2^K$ ground states has $S^z=-1$ for all spins on the inner
ring while the spins on the outer ring are arbitrary.
By removing the $Z$ direction field from one of the spins on the inner ring, the model (\ref{toy}) has a unique
 unique ground state with $S^z_i=+1$ for all $i$ while the cluster of states with $S^z=-1$ on the inner ring is now an excited state with energy $1/2$
above the ground state.

Now consider the effect of a small transverse magnetic field as occurs near the end of an annealing path.
The energy of the unique ground state does not change to linear order in the transverse field strength.  However, the energy of
the cluster of states does change to linear order, by an amount proportional to the number of spins.
Thus
such a low order perturbation analysis suggests a level crossing occuring at a transverse magnetic field strength proportional to $1/N$, i.e., a level
crossing in $H_s$ for $(1-s) \sim 1/N$.
Of course, since $H_s$ always has a unique ground state this level crossing must become an avoided crossing.  However, $K\sim N$ spins must flip to move
from the cluster to the core, so one may expect that the gap will be small, proportional to the transverse magnetic field strength raised to a power proportional to $K$.  Thus, the gap will be of order $N^{-{\rm const}\times N}$ for some positive constant.

This argument for the small gap above is closely related to the argument of Ref.~\onlinecite{MBL}, and so this toy model may provide an interesting
example.  It would be interesting if a super-exponentially small gap could be proven in this particular case.

The performance of various schedules in this model is shown in table \ref{tabletoy}.
For $K=2$, the slow annealing schedule $L(80,1,1)$ outperforms the others, but already its success probability is noticeably less than $1$.  For $K=3$, the slow anneal $L(80,1,1)$ and the fast anneal $L(10,1,1)$ have comparable performance, and for increasing values of $K$, the slow anneal becomes dramatically worse.  This is due to the spectrum of the model which has a single avoided crossing with very small gap.  Comparing $L(10,1,1)$ to $154$, we find that $154$ is consistently better and becomes relatively better as $K$ increases.  Both $L(10,1,1)$ and $154$ show a roughly exponential decay of the squared overlap with increasing $K$, but the decay is slightly faster for $L(10,1,1)$.

\begin{table}
\label{tabletoy}
\begin{tabular}{c|c|c|c}
K & 154 & L(10,1,1) & L(80,1,1) \\
2 &0.409 &0.379 &0.811\\
3 &0.237 &0.208& 0.212\\
4 &0.157 &0.104 &0.0182\\
5 &0.1   &0.0493 &0.000683\\
6 &0.0582 &0.0233 &1.25$\times 10^{-5}$\\
7 &0.0313 &0.011 &9.37$\times 10^{-6}$\\
8 &0.0169 &0.00524 &1.34$\times 10^{-5}$\\
9 &0.0095 &0.00248 &4.$\times 10^{-6}$\\
10& 0.00543 &.00118 &5.42$\times 10^{-7}$
\end{tabular}
\caption{Absolute squared overlap for various values of $K$, for learned schedule $154$ and for annealing schedules $L(10,1,1)$ and $L(80,1,1)$.}
\end{table}

We remark that we removed the $Z$ field from one of the inner spins to break the ground state degeneracy.  Another way to do this is to
vary the field strengths, keeping the same field on all inner spins but making it slightly weaker.  The results are shown in Table \ref{tabletoy2}, where we took the inner field strength to be $(1/4)(N-1)/N$ on all spins (so that the total field is the same as above).  It is interesting that this does not hurt the performance of the learned schedule (see discussion of weighted MAX-2-SAT later).

\begin{table}
\label{tabletoy2}
\begin{tabular}{c|c|c|c}
K & 154 & L(10,1,1) & L(80,1,1) \\
2 & .422 & .386 & 0.8\\
3 & .265 & .228 & .191\\
4 & .186 & .122 & .0124\\
5 & .121 & .0594 & .000353\\
6 & 0.0704 & 0.0283 & .000214\\
7 & 0.0379 & .0135 & .000113\\
8 & .0204 & .00647 & 3.98$\times 10^{-5}$\\
9 & .0115 & .00309 & 6.79$\times 10^{-6}$\\
10 & .0066 & .00147 & 2.15$\times 10^{-7}$
\end{tabular}
\caption{Absolute squared overlap for various values of $K$, for learned schedule $154$ and for annealing schedules $L(10,1,1)$ and $L(80,1,1)$.  All inner fields have same strength but are reduced compared to outer fields.  Total strength of inner fields is same as in Table \ref{tabletoy}.}
\end{table}

We also studied another toy model.  This model has $N$ states (not $N$ qubits, but rather an $N$ dimensional Hilbert space), divided into $3$ subspaces of dimensions $N_1,N_2,1$ respectively, with $N=N_1+N_2+1$.  The Hamiltonian $H_X$ was chosen to be a sum of two terms; the first term was proportional to the projector onto the uniform superposition of all states, while the second term was proportional to the uniform projector onto the superposition of states in the second subspace (the subspace with dimension $N_2$).  The Hamiltonian $H_Z$ was proportional to the identity in each eigenspace, with the ground state being the third subspace (of dimension $1$) and the subspace of first excited states being the subspace of dimension $N_2$.  If we take $N_2=0$, then this model is simply an instance of database search (and Grover's algorithm is optimal\cite{grover}); likely there are algorithms similar to Grover which are equally optimal for this model.  However, our goal was instead to test various schedules.  We found that if the second term in $H_X$ was chosen sufficiently strong, then this would create a small gap: at intermediate values of $s$ the ground state was concentrated on the second subspace while at $s=1$ the ground state was the third subspace.
It was in this case that the learned schedules outperformed the annealing schedules.

\subsection{Creating Excited States}
These toy models suggest the following explanation for the success of the learned schedules.  Small gaps can create difficulties for an annealing algorithm.  These small gaps can occur especially if one ``basin of local minima" has slightly higher energy than the true minimum of $H_Z$ but is able to reduce its energy by more in the presence of a transverse field.
Suppose there is a single small gap at some $s_c$, with the gap sufficiently small that a very slow anneal will be required to stay in the ground state.
In this case, it might be desirable to be in an {\it excited} state at an intermediate value of $s$ ($s<s_c$) and then to anneal more rapidly so that one ends close to the ground state for $s>s_c$.

There are a variety of possible ways to produce this excited state.  In Ref. \onlinecite{dwave1}, thermal excitation was suggested as one possible mechanism.  The optimized anneals of CFLLS give another mechanism.  Let us say that $T_{slow}$ is some characteristic timescale to stay in the ground state for $s$ near $s_c$, while $T_{int}$ is some intermediate timescale required to stay in the ground state for other values of $s$.
Thus, a fast anneal (faster than $T_{int}$) may lead to a transition to an excited state at some small $s$, leaving one in the appropriate excited state at $s$ slightly smaller than $s_c$.

Another strategy also tried in CFLLS was to deliberately prepare the system in a {\it randomly chosen} first excited state at $s=0$ and then run an anneal (the time of this anneal might be longer than $T_{int}$ but still faster than $T_{slow}$) so that one is hopefully in the first excited state at $s$ slightly smaller than $s_c$.  Note that there are $N$ degenerate first excited states at $s=0$ so the probability of success of this method is at most $1/N$.  It was found\cite{EC} that in fact the probability of success was close to $1/N$.

However, the learned schedules in this paper give a higher probability of success than this (significantly higher than $1/N$ for most of the instances).  Thus, we conjecture that the success of the learned schedules is that the behavior in the first steps (with an oscillating $Z$ term, and a large $X$) serve to drive the system into the {\it correct} first excited state and then schedules conclude by approximately following an anneal so that they end in the ground state.

\section{Discussion}
We have applied a numerical search to find schedules for a modification of the QAOA algorithm.  These schedules were trained on a small subset of instances with $20$ bits, but were found to perform well on the full set of such instances as well as related but slightly different ensembles with $20,24$ and $28$ bits.
The performance of these schedules raises the hope that they may outperform annealing on larger sizes and may be a useful application for
an early quantum computer.

As a caveat, we have only studied SAT problems.  We began a study of {\it weighted} SAT, where each clause comes with some arbitrary energy cost for violating that clause.  As a first step to such a study, we simply tried giving all clauses the same weight; this does not change the ground state of $H_Z$ but simply scales $H_Z$ by some factor.  However, the learned schedules did not perform well even with this simple rescaling.  By training the schedules instead on a range of such weighted instances (for example, training on a set of $10$ random instances as well as those instances rescaled by various factors) we were able to slightly improve the ability to deal with this rescaling, but the ratios were much worse than the results reported here.
It may be the case that other initial schedules or training methods would better deal with this case.

For hardware implementation, we have studied some schedules where $\theta^Z$ simply does a linear ramp, which may be easier to implement.
Further, any schedule where $\theta^Z$ has a fixed sign can be implemented by taking a time-varying $\theta^X$ and a time-constant $\theta^Z$.  That is, suppose one has the abiility to time-evolve under the Hamiltonian $g^X H_X + g^Z H_Z$ for arbitrary $g^X$ and some given $g^Z$; then, to
implement a unitary transformation $\exp[i (\theta^X H_X + \theta^Z H_Z)]$ one should evolve under the Hamiltonian $g^X H_X+g^Z H_Z$ for
$g^X=g^Z \theta^X/\theta^Z$ and do the evolution
for time $\theta^Z/g^Z$.

We have found that it is very important to have an appropriate initial schedule as otherwise the learning gets trapped in local optima.
Thus, while it may be the case that one can learn a schedule on a classical computer using a modest number of qubits and then apply it on a quantum
computer with a larger number of qubits, the learned schedule might also be a good starting point for further optimization of schedules on the quantum computer.

\acknowledgements
We thank E. Crosson for supplying the instances in CFLLS and for very useful explanations.  We thank E. Farhi for useful comments on a draft of this paper.
This work was supported by Microsoft Research. We acknowledge hospitality of the Aspen Center for Physics, supported by NSF grant PHY-1066293.

\onecolumngrid
\appendix

\section{Schedules}
Here we give the parameters for certain learned schedules.

\begin{table}[h]
\begin{tabular}{c|c|c|c|c|c|c|c|c|c|c|c|}
Schedule & Initial & $\theta^Z_1$ & $\theta^Z_2$ & $\theta^Z_3$ & $\theta^Z_4$ & $\theta^Z_5$ & $\theta^Z_6$ & $\theta^Z_7$ & $\theta^Z_8$ & $\theta^Z_9$ & $\theta^Z_{10}$ \\
\hline
8 & 8 & -0.279307 & 0.313947 & 0.614148 & -0.220295 & 0.256869 & 0.465194 & -0.212299 & 0.312254 & 1.50651 & 2.011013 \\
31 & 9 & 0.368606 & 0.359748 & 0.190667 & 0.392364 & 0.208514 & 0.021365 & 0.642995 & 1.143198 & 1.64574 & 1.814225\\
49 & 9 & 0.424251 & 0.771576 & 0.464935 & 0.435078 & 0.404496 & 0.187802 & 0.77197 & 1.300528 & 1.701031 & 1.745732\\ 
84 & 11 & 0.1629 & -0.496857 & 0.450711 & -0.791892 & 0.326329 & -0.475372 & 0.433593 & 1.033271 & 1.659841 & 2.031027\\
113 & 12 & 0.37599 & 0.680923 & 0.997025 & 0.715514 & 0.271968 & 0.519316 & 1.068852 & 1.443309 & 1.433469 & 1.333607\\
122 & 12 & 0.489956 & 0.510331 & 0.740654 & 0.538733 & 0.245925 & 0.08665 & 0.761729 & 1.188631 & 1.418336 & 1.89151\\
154 & 14 & 0.748224 & -0.080047 & -0.117857 & 0.316126 & 0.096738 & -0.307805 & 1.210155 & 1.183015 & 1.557269 & 1.745549\\
157 & 14 & 0.677717 & -0.099922 & -0.055678 & 0.294502 & 0.107643 & -0.276445 & 1.070014 & 1.057304 & 1.479656 & 1.646192
\end{tabular}
\caption{$\theta^Z$ for certain learned schedules.  First column gives key indicating particular learned schedule number (the number itself is meaningless and serves only as a key.
Second column gives initial schedule for training (see table \ref{tabsched}).}
\end{table}
\begin{table}[h]
\begin{tabular}{c|c|c|c|c|c|c|c|c|c|c|c|}
Schedule & Initial & $\theta^X_1$ & $\theta^X_2$ & $\theta^X_3$ & $\theta^X_4$ & $\theta^X_5$ & $\theta^X_6$ & $\theta^X_7$ & $\theta^X_8$ & $\theta^X_9$ & $\theta^X_{10}$ \\
\hline
8 & 8 & 0.985164 & 1.711707 & 1.308381 & 1.272364 & 0.71373 & 2.073916 & 1.340572 & 1.037615 & 1.217506 & 0.730447 \\
31 & 9 & 1.168114 & 1.375238 & 1.350988 & 1.356165 & 1.337642 & 1.091975 & 1.426565 & 1.162721 & 0.885662 & 0.431466 \\ 
49 & 9 & 1.510793 & 1.665954 & 1.205267 & 1.062189 & 1.59617 & 1.481757 & 1.6141 & 1.285973 & 0.903954 & 0.396039 \\
84 & 11 & 1.945308 & 1.142874 & 0.875239 & 0.914909 & 1.373274 & 1.191093 & 2.016909 & 1.142808 & 1.104454 & 0.585 \\
113 & 12 & 1.609044 & 1.459435 & 1.971842 & 1.625206 & 1.537716 & 1.515011 & 1.398038 & 0.983823 & 0.5701 & 0.273691 \\ 
122 & 12 & 1.683547 & 0.979162 & 1.878078 & 1.631202 & 1.16941 & 1.055429 & 1.635904 & 1.172053 & 0.795996 & 0.519226 \\
154 & 14 & 1.35801 & 0.955197 & 1.397257 & 1.219015 & 1.396977 & 1.420552 & 1.283791 & 0.889047 & 0.671747 & 0.339493 \\
157 & 14 & 1.359167 & 1.060199 & 1.293059 & 1.248988 & 1.328482 & 1.431533 & 1.237331 & 0.854213 & 0.688784 & 0.382808 
\end{tabular}
\caption{$\theta^X$ for certain learned schedules.  First column gives key indicating particular learned schedule number (the number itself is meaningless and serves only as a key.
Second column gives initial schedule for training (see table \ref{tabsched}).}
\end{table}

\begin{thebibliography}{99}
\bibitem{detail} In detail, in a random step, each parameter is incremented by a value chosen independently and uniformly from the interval $[-\delta,+\delta]$, where initially $\delta=0.1$.  If fewer than $3$ trials improve the objective out of $50$ attempts, then $\delta$ is multiplied by $1/2$, while if more than $5$ improve it, $\delta$ is multiplied by $2$.
\end{thebibliography}
\end{document}